\documentclass{article}

\usepackage{arxiv}

\usepackage[utf8]{inputenc} 
\usepackage[T1]{fontenc}    
\usepackage{hyperref}       
\usepackage{url}            
\usepackage{booktabs}       
\usepackage{amsfonts}       
\usepackage{nicefrac}       
\usepackage{microtype}      
\usepackage{lipsum}
\usepackage{color}
\usepackage[round]{natbib}
\usepackage{graphicx}

\title{Multi-modal Emotion Detection \\ with Transfer Learning}

\author{
  Amith Ananthram$^1$ \\
  \And
  Kailash Karthik Saravanakumar$^1$ \\
  \And
  Jessica Huynh$^1$ \\
  \And
  Homayoon Beigi$^{1,2}$ \\
  \And
 \\
 \begin{minipage}{0.48\textwidth}
  \centering
   \textsuperscript{1}Columbia University \\
   New York, NY 10027
 \end{minipage}
 \begin{minipage}{0.48\textwidth}
  \centering
   \textsuperscript{2}Recognition Technologies, Inc. \\
   South Salem, NY 10590
 \end{minipage}
 }

\begin{document}
\maketitle

\begin{abstract}
Automated emotion detection in speech is a challenging task due to the complex interdependence between words and the manner in which they are spoken. It is made more difficult by the available datasets; their small size and incompatible labeling idiosyncrasies make it hard to build generalizable emotion detection systems. To address these two challenges, we present a multi-modal approach that first transfers learning from related tasks in speech and text to produce robust neural embeddings and then uses these embeddings to train a pLDA classifier that is able to adapt to previously unseen emotions and domains. We begin by training a multilayer TDNN on the task of speaker identification with the VoxCeleb corpora and then fine-tune it on the task of emotion identification with the Crema-D corpus.  Using this network, we extract speech embeddings for Crema-D from each of its layers, generate and concatenate text embeddings for the accompanying transcripts using a fine-tuned BERT model and then train an LDA - pLDA classifier on the resulting dense representations. We exhaustively evaluate the predictive power of every component: the TDNN alone, speech embeddings from each of its layers alone, text embeddings alone and every combination thereof.  Our best variant, trained on only VoxCeleb and Crema-D and evaluated on IEMOCAP, achieves an EER of $38.05$\%. Including a portion of IEMOCAP during training produces a 5-fold averaged EER of $25.72$\% (For comparison, $44.71$\% of the gold-label annotations include at least one annotator who disagrees).
\end{abstract}

\keywords{Emotion Detection, Multimodal Embeddings, Transfer Learning, Domain Adaptation, pLDA}

\section{Introduction}
Due to the growing presence of AI-powered systems in our lives, affective computing has become an important part of human-computer interaction. Emotion plays a role in our thoughts and actions and is an integral part of the way we communicate \citep{conv_atten_emo}. The ability to leverage context to understand emotions communicated both verbally and non-verbally is trivial for humans but remains difficult for machines \citep{compl_fusion}. Emotional responses depend on both our psyche and physiology and are governed by our perception of situations, people and objects. They also depend on our mental state (mood, motivation, temperament)  \citep{beigi_iemocap}. The way we exhibit and perceive emotion may also differ based on our age, gender, race, culture and accent \citep{1904.03833}. In addition to all of this, unlike targets in other classification tasks, the emotions we experience are rarely distinct: they often coexist without clear temporal boundaries, adding considerable complexity to the task \citep{1704.08619}.

Despite these difficulties, automated emotion recognition has social and commercial applications that make it worth pursuing. In the medical domain, it has exciting potential: to identify and diagnose depression and stress in individuals \citep{depression, distress}, to monitor and help people with bipolar disorder \citep{bipolar} and to assist the general public in maintaining mental health. Commercial applications include call center customer management, advertising through neuro-marketing and social media engagement \citep{1704.08619, conv_atten_emo, compl_fusion}. As intelligent chatbots and virtual assistants have become more widely used, emotion detection has become a vital component in the design, development and deployment of these conversational agents \citep{1810.04635}.

Early research in emotion detection focused on binary classification in a single modality, whether in text, speech \citep{speech_1, speech_2}, or images \citep{image_1}. Text-based classifiers used the n-gram vocabulary of sentences to predict their polarity and speech models modeled the vocal dynamics that characterize these emotions. These approaches are inherently limited: a binary granularity and cues from a single modality are far removed from the actual human process they're meant to model. As a result, joint approaches which leverage all available modalities (e.g., both speech and text in applications like home assistants) are promising.

\begin{figure}[t]
    \centering
    \includegraphics[width=\linewidth]{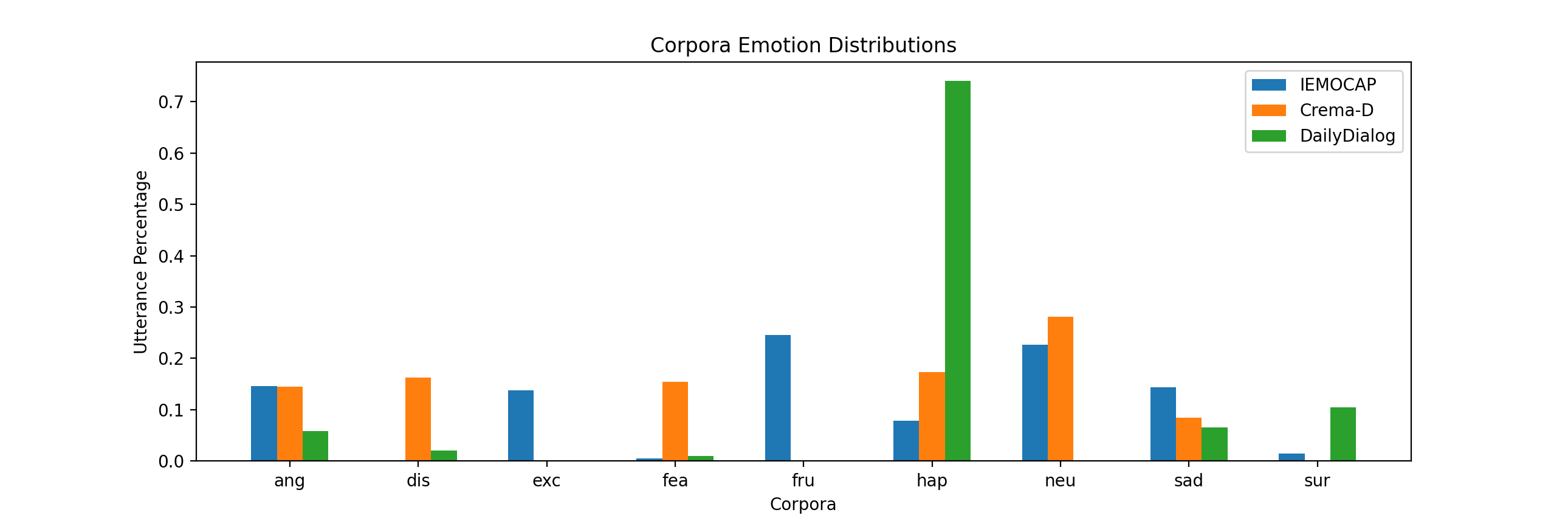}
    \caption{Emotion label distributions in our training and evaluation corpora.}
    \label{image:dists}
\end{figure}

While existing multi-modal emotion corpora like IEMOCAP \citep{iemocap} and Crem-D \citep{cremad} have been critical for the progress in affective computing to date, they suffer from three issues that are the focus of our work. First, these corpora tend to be small due to the high costs of annotating for emotion. This precludes the use of deep neural models with high model complexity as they require many training samples to generalize well. This also compounds the second difficulty inherent to many emotion datasets: while there are usually many neutral, happy and sad training examples, there are often very few examples of rarer emotions like disgust making them difficult to classify. This issue is not easily solved by combining different corpora due to the third issue, their lack of mutual compatibility -- they differ in the emotions identified, the types of dialogue and number of speakers represented and the naturalness of the recordings (see Figure \ref{image:dists}). This severely restricts the generalizability of models trained on a single corpus.

Contemporary literature has dealt with these problems by dropping labels (\cite{pappagari2020x, chen2020multi, yoon2020attentive}). Hard and scarce emotions like disgust are dropped from the corpus and the models are trained and evaluated on the trimmed corpus. This allows evaluating models on different corpora by using utterances exhibiting only the most common emotions. While this is a reasonable, the resulting performance is not a complete reflection of how these models perform once deployed to production. When emotion models are used in real-world applications, we can expect them to encounter utterances corresponding to dropped labels. For such cases, these models are likely to exhibit degraded performance by predicting one of the known, but incorrect labels.

In this work, we address the problem of data sparsity by transfer learning via the pretrain-then-finetune paradigm. Deep complex models can be trained on large datasets for an auxiliary but related task to learn network parameters that reflect abstract notions related to the target task. As the expression of emotions is highly dependent on the individual, we train a multilayer TDNN (\cite{waibel1989phoneme}) on the task of speaker identification using the VoxCeleb corpus (\cite{voxceleb}) and then fine-tune its final few layers on the task of emotion identification using the Crema-D corpus (\cite{cremad}). Using this network, we extract speech embeddings for Crema-D from each of its layers, generate and concatenate text embeddings for the accompanying transcripts using a fine-tuned BERT model (\cite{bert}) and then train an LDA - pLDA (\cite{fisher1936use, ioffe2006probabilistic}) model on the resulting dense representations.  pLDA allows our model to more easily adapt to previously unseen classes and domains, a requirement for both evaluating against a different emotion corpus with an incompatible label set and performing well in the wild.

To understand the merits of each component, we exhaustively evaluate the predictive power of every permutation: the TDNN alone, speech embeddings from each of its layers alone, text embeddings alone and every combination thereof. Our best variant, trained on only VoxCeleb and Crema-D and evaluated on IEMOCAP, achieves an Equal Error Rate (EER) of $38.05$\%. Including a portion of IEMOCAP during training produces a 5-fold averaged EER of $25.72$\%. 

\section{Related Work}

\subsection{Problem Formulation}

In this work, we focus on two tasks related to emotions.  The first is \textit{emotion identification}.  Given the audio and accompanying text for an utterance $\mathbf{u}$, \textit{emotion identification} is the task of identifying the emotion $\mathbf{e}$ expressed in $\mathbf{u}$ from a fixed set of emotions $\mathbf{E}$.  This is the standard classification task found in the literature.

The second is \textit{emotion confirmation}.  Given the audio and accompanying texts for two utterances $\mathbf{u_1}$ and $\mathbf{u_2}$, \textit{emotion confirmation} is the task of identifying whether the two utterances express the same emotion.  This task can be thought of as analogous to either hypothesis testing in speaker recognition or a one-shot classification task.  It is motivated by the labeling mismatches among the various emotion corpora and is meant to better reflect the requirement that emotion detection systems be able to adapt to emotions unseen during training once deployed to production.

\subsection{Unimodal Speech Emotion Detection}

Early work on emotion detection in speech focused on the extraction of hand-crafted features for classification. \citet{liscombe_features} extracted a set of continuous features based on the fundamental frequency, amplitude and spectral tilt of speech and analyzed its correlation with different emotions. Contemporary literature has focused on deep neural networks with particular successes in transfer leraning. \citet{pappagari2020x} studied the transfer of embeddings from a ResNet-based speaker verification model using linear methods. Transfer learning with TDNNs has also been shown to be effective. \citet{zhou2020transfer} fine-tune a pre-trained ASR model and achieve strong results. An alternate approach seen in the literature is to train in a multi-task context on a useful auxiliary task \citep{goel2020cross}.

\subsection{Multi-modal Emotion Detection}

Deep neural techniques have often been applied to integrate information from both the speech and text modalities. \citet{heusser2019bimodal} trained separate speech and text emotion classifiers and then jointly optimized them for multimodal emotion detection. An alternate method for creating multimodal classifiers is to use the embeddings from a hidden layer of the unimodal models for multimodal analysis. While \citet{chen2020multi} created an ensemble of classifiers using the unimodal embeddings individually and their multimodal concatenations, \citet{tripathi2019deep} fed the concatenation to a fully-connected neural layer and backpropagated the classification loss. Fusion using attention is yet another method for combining embeddings from different modalities \citep{lian2019domain}. Even in the trimodal setting (speech, text and vision), the primary approaches for integrating modalities are concatenation and attentive fusion \citep{yoon2020attentive, mittal2019m3er, tripathi2018multi}.

Contemporary results in emotion detection on IEMOCAP are shown in Table \ref{table:related_work} with their accompanying modalities, supported emotion sets and performances.

\begin{table}[t]
\centering
\caption{Contemporary emotion detection models, with their modalities, emotion sets and average IEMOCAP \citep{iemocap} performance.  These models were trained on $4/5$ of IEMOCAP and evaluated on the held out fifth.}
\begin{tabular}{lllc}
\toprule
\multicolumn{1}{c}{\textbf{Model}} & \multicolumn{1}{c}{\textbf{Emotion Labels}} & \multicolumn{1}{c}{\textbf{Metrics}} & \multicolumn{1}{c}{\textbf{Performance}} \\
\midrule
\multicolumn{4}{c}{\emph{Unimodal - Speech}} \\
\midrule
\citet{pappagari2020x} & happy+exc, angry, sad, neutral & Weighted F1 & $0.70$ \\
\citet{goel2020cross} & happy, angry, sad, neutral and fear & Accuracy & $0.51$ \\
\citet{zhou2020transfer} & happy+exc, angry, sad, neutral & Accuracy & $0.72$ \\
\midrule
\multicolumn{4}{c}{\emph{Bimodal - Speech and Text}} \\
\midrule
\citet{chen2020multi} & happy, angry, sad, neutral & \begin{tabular}[l]{@{}l@{}}Weighted Accuracy \\ Unweighted Accuracy\end{tabular} & \begin{tabular}[c]{@{}l@{}}$0.71$ \\ $0.72$
\end{tabular} \\
\citet{heusser2019bimodal} & happy, angry, sad, neutral & \begin{tabular}[l]{@{}l@{}}Weighted Accuracy\\ Unweighted Accuracy\end{tabular} & \begin{tabular}[c]{@{}l@{}}$0.70$\\ $0.67$\end{tabular} \\
\citet{lian2019domain} & happy+exc, angry, sad, neutral & Weighted Accuracy & $0.83$ \\
\citet{tripathi2019deep} & happy, angry, sad, neutral & \begin{tabular}[c]{@{}l@{}}Accuracy\\ Class Accuracy\end{tabular} & \begin{tabular}[c]{@{}l@{}}$0.76$\\$0.69$\end{tabular} \\
\midrule
\multicolumn{4}{c}{\emph{Trimodal - Speech, Text and Vision}} \\
\midrule
\citet{yoon2020attentive} & \begin{tabular}[l]{@{}l@{}}happy, angry, sad, neutral \\ excited, frustrated, surprised\end{tabular} & \begin{tabular}[l]{@{}l@{}}Weighted Accuracy \\ Unweighted Accuracy\end{tabular} & \begin{tabular}[c]{@{}l@{}}$0.62$ \\ $0.60$\end{tabular} \\
\citet{mittal2019m3er} & happy, angry, sad, neutral & \begin{tabular}[l]{@{}l@{}} Mean Accuracy\\ F1\end{tabular} & \begin{tabular}[c]{@{}l@{}} $0.82$\\ $0.82$\end{tabular} \\
\citet{tripathi2018multi} & happy+exc, angry, sad, neutral & Accuracy  & $0.71$ \\
\bottomrule
\end{tabular}
\label{table:related_work}
\end{table}

\section{Data}

\begin{table}
\centering
\caption{Mapping of Emotions from Corpus Labels to Canonical Classes}
\begin{tabular}[t]{lcccc}
\toprule
\textbf{Canonical Emotion} & \textbf{IEMOCAP} & \textbf{Crema-D} & \textbf{DailyDialog}\\
\midrule
\vspace{0.1cm}
Happiness & Happiness, Excitement & Happiness, Excitement & Happiness\\
\vspace{0.1cm}
Sadness& Sadness & Sadness & Sadness\\
\vspace{0.1cm}
Fear/Surprise & Fear, Surprise & Fear & Fear, Surprise\\
\vspace{0.1cm}
Anger/Disgust & Anger, Disgust, Frustration & Anger, Disgust & Anger, Disgust\\
\vspace{0.1cm}
Neutral & Neutral & Neutral & Other\\
\bottomrule
\end{tabular}
\label{table:emotion_classification}
\end{table} 

\subsection{Emotion Classes}

As previously discussed, the set of emotions labeled in different corpora is not uniform.  A classic model of emotions from psychology comes from Ekman \citep{ekman} who identifies six fundamental emotions -  happiness, sadness, anger, disgust, surprise and fear. Computational models of emotions have attempted classification in this six emotion setting. These approaches have struggled with anger, disgust and fear as they are usually underrepresented in training corpora \citep{meld}. As a result, literature has typically resorted to label dropping, reporting substantially improved performance on the resulting trimmed corpus. 

Contemporary research in psychology, informed by facial dynamics, has proposed further clustering into four basic emotions \citep{four_emotions}, arguing against the six-emotion theory. The authors suggest that two pairs of emotions - (\textit{anger, disgust}) and (\textit{fear, surprise}) are psychologically irreducible, a possible explanation for the performance drop seen in systems attempting to disambiguate between these pairs.  In light of this research, we adopt their suggested grouping to better balance the classes in our training corpus, Crema-D.  When evaluating our fine-tuned neural model on the task of \textit{emotion identification} in IEMOCAP, we rely on this grouping again.  While we do the same when evaluating our pLDA classifier on the task of \textit{emotion confirmation}, we also present its performance on IEMOCAP's ungrouped emotions, testing its ability to generalize to previously unseen classes, a key focus of this work.  When grouping utterances, we combine the indistinguishable classes into the canonical classes shown in Table \ref{table:emotion_classification}.

 \subsection{Pre-Training Corpora for Auxiliary Tasks}
 
 \subsubsection{VoxCeleb1 \& VoxCeleb2, for Speaker Recognition}

 Our speech model is pre-trained on the auxiliary task of speaker recognition (\cite{r:beigi-sr-book-2011}) to learn embeddings that model basic vocal characteristics. VoxCeleb1 and VoxCeleb2  \citep{voxceleb} are natural speech datasets with audio and video clips of nearly 8,000 celebrities uploaded to YouTube. In total, the two datasets contain more than 1 million utterances. They are fairly well-balanced and the speakers included are reasonably diverse (66\% male with many different ethnic groups and nations represented). These recordings are not acted and include background noise.

\subsection{Emotion Fine-Tuning and Evaluation Corpora}

\subsubsection{Crema-D}
 
Crema-D is a multi-modal emotion dataset in the audio and visual modalities (\cite{cremad}). Actors perform a limited set of 12 utterances in every target emotion (masking the relationship between their semantic content and their emotional thrust). The emotions labeled are anger, disgust, excitement, fear, happiness, neutral and sadness and include an associated intensity rating (low, medium, high or unspecified). The dataset contains 7,442 utterances from 91 actors who are chosen across different age and ethnic groups.  We use Crema-D as our emotion fine-tuning corpus.

\subsection{DailyDialog}

While Crema-D is a multi-modal emotion dataset, the lack of diversity in the content of the utterances makes it ineffective to train a text-based emotion recognition model. To address this, we use DailyDialog (\cite{li-etal-2017-dailydialog}), a multi-turn dialog dataset with emotion annotations. The dataset contains more than 13,000 dialogues corresponding to over 100,000 utterances. The emotions labeled are anger, disgust, fear, happiness, sadness, surprise and "other". 
 
\subsubsection{IEMOCAP}
 
Interactive Emotional Dyadic Motion Capture (IEMOCAP) \citep{iemocap} consists of 12 hours of audio and video clips performed by 5 male and 5 female actors. These recordings include improvisations and scripted conversations, both designed to elicit certain emotions. They are then split into individual utterances and annotated for both dimensional attributes and categorical attributes: anger, disgust, excitement, fear, frustration, happiness, neutral, sadness and surprise. Utterances for which human annotator agreement could not be reached are labelled \textit{xxx} and the fraction of the dataset with this label is around 25\%. As most emotion recognition systems evaluate against IEMOCAP, we do as well in two contexts: without including any IEMOCAP during training and with $4$ out of $5$ sessions included during training with the fifth held out for evaluation as part of a five-fold cross validation.
 
\section{Methodology}

\begin{figure}[t]
    \centering
    \includegraphics[width=0.5\linewidth]{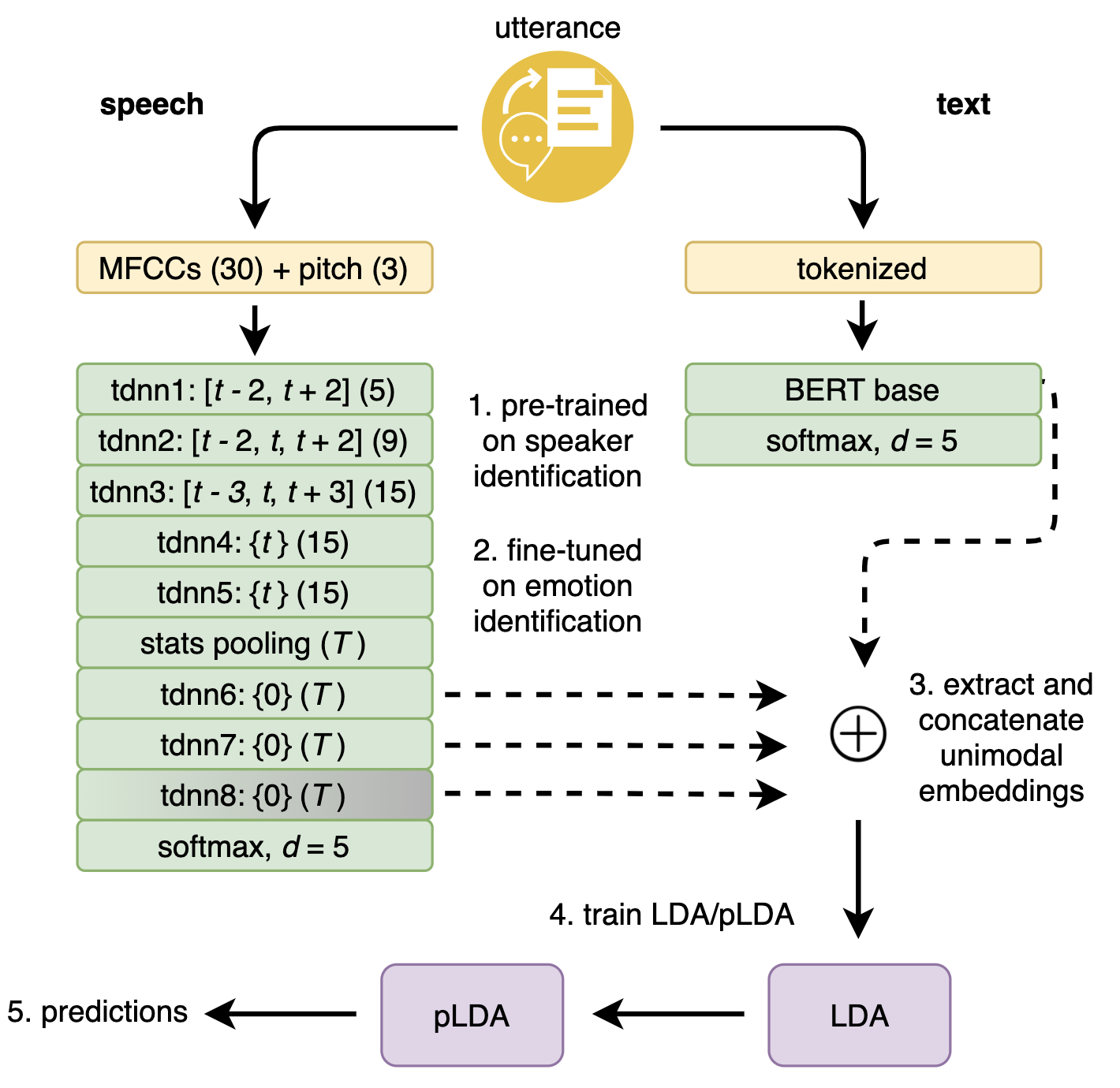}
    \caption{Architecture of our multi-modal emotion detection pipeline}
    \label{image:arch}
\end{figure}

Our emotion detection pipeline is comprised of the following three stages: (1) extraction of unimodal embeddings; (2) concatenation of unimodal embeddings to create a single multimodal embedding; and (3) inference on the multimodal embedding vector with an LDA/pLDA classifier. This architecture is depicted in Figure \ref{image:arch}. 

\subsection{Unimodal Neural Networks}

Our embedding networks follow the pretrain-then-finetune training paradigm.

Our speech embedding model is a time-delay neural network (TDNN, \cite{waibel1989phoneme}).  TDNNs are designed to capture long term temporal dependencies efficiently; lower layers aggregate information within narrow contexts while higher layers learn more abstract representations.  In \citet{xvectors}, the authors show that training a pLDA classifier on fixed-length embeddings extracted from the higher layers of a speaker recognition TDNN (which they refer to as "x-vectors") achieves superior performance on out-of-class speaker recognition.  Inspired by their result, we choose the same auxiliary task as we hypothesize that such a network learns dense representations of speech segments in its upper layers that contain signals relevant to emotion detection too.  We extract hand-crafted speech features and pre-train a TDNN on speaker identification (\cite{r:beigi-sr-book-2011}). We adopt their architecture (described in Table \ref{table:spkr_model_arch}) and their training methodology, building on its accompanying training script published as part of the Kaldi toolkit (\cite{povey2011kaldi}) with a slight modification. We include $3$ pitch features, hypothesizing that these contain signal useful for our eventual fine-tuning on emotion identification.  

After pre-training on speaker identification, we fine-tune this model on the task of emotion identification using the canonical labelings identified in 3.1.  We extract the same set of features and fine-tune several variants, experimenting with learning rates for the first six layers, adding an eighth layer and augmenting our emotion corpus with noise (which has proven to produce more robust embeddings for speaker recognition, \cite{xvectors}) .

For embedding the audio transcript text, we fine-tune BERT on the task of emotion identification. The resulting speech and text embeddings are then concatenated to produce a multimodal embedding vector. 

\subsection{LDA/pLDA}

The final stage in our pipeline is an LDA/pLDA \citep{ioffe2006probabilistic, r:beigi-sr-book-2011} classifier.  Linear discriminant analysis, or LDA, is a dimensionality reduction technique designed to find a set of linear features that maximize the between-class separation of data while minimizing the within-class
scatter.  Probabilistic linear discriminant analysis, or pLDA, is a generative model that associates probability distributions with those linear features.  This makes it adaptable to classes unseen during training, enabling both \textit{emotion identification} and \textit{emotion confirmation}.  We train this classifier on our multimodal embeddings using the canonical labelings identified in 3.1.

Experimental details for training each component in this pipeline can be found in Section 5.

\section{Experiments}

\subsection{Experimental Setup}

\subsubsection{Data Preprocessing}

We extract hand-engineered speech features for training our speaker recognition model in the Kaldi toolkit (\cite{povey2011kaldi}): the top $30$ MFCCs (\cite{r:beigi-sr-book-2011}) and $3$ additional pitch features (probability of voicing, mean-subtracted log, and raw pitch deltas) over a frame-length of 25ms. These features are then normalized with sliding-window cepstral mean normalization with a window of $300$ frames.

\subsubsection{Pre-Training on Speaker Recognition}

Using the methodology presented in 4.1, we pre-train our TDNN (Table \ref{table:spkr_model_arch}) on the task of speaker identification using the VoxCeleb1 and VoxCeleb2 corpora.  We adopt the hyperparameters used by \citet{xvectors} in their entirety.

\begin{table}
\centering
\caption{Speaker recognition model architecture, where $N$ is the number of speakers in the training corpus}
\begin{tabular}[t]{lcc}
\toprule
\textbf{Layer (Context)} & \textbf{Layer Context} & \textbf{Input x Output}\\
\midrule
tdnn1 & [$t - 2$, $t + 2$] ($5$) & $165$ x $512$ \\
\vspace{0.1cm}
tdnn2 & \{$t - 2$, $t$, $t + 2$\} ($9$) & $1536$ x $512$ \\
\vspace{0.1cm}
tdnn3 & \{$t - 3$, $t$, $t + 3$\} ($15$) & $1536$ x $512$ \\
\vspace{0.1cm}
tdnn4 & \{$t$\} ($15$) & $512$ x $512$ \\
\vspace{0.1cm}
tdnn5 & \{$t$\} ($15$) & $512$ x $1500$ \\
\vspace{0.1cm}
stats pooling & [$0$, $T$) ($T$) & $1500T$ x $3000$ \\
\vspace{0.1cm}
tdnn6 & \{$0$\} ($T$) & $3000$ x $512$ \\
\vspace{0.1cm}
tdnn7 & \{$0$\} ($T$) & $512$ x $512$ \\
\vspace{0.1cm}
softmax & \{$0$\} ($T$) & $512$ x $N$ \\
\bottomrule
\end{tabular}
\label{table:spkr_model_arch}
\end{table}

\subsubsection{Fine-Tuning Speech on Emotion Detection}

We then fine-tune our speaker recognition model on the task of emotion detection using the Crema-D corpus with the canonical label clustering described in Table 3.1.  Employing the methodology described in 4.1, we optimize a cross entropy loss using stochastic gradient descent with momentum.  We use an initial learning rate of $1e^{-3}$, a final learning rate of $1e^{-4}$, a batch size of $64$, a dropout rate of $50\%$ and we train for $3$ epochs.  We fine-tune several variants by varying the learning rates on the first six layers, adding an eighth layer and augmenting our emotion corpus with noise.

\subsubsection{Fine-Tuning Text on Emotion Detection}

We use the uncased version of BERT base (twelve 768 dimensional layers) as our pre-trained text embedding model\footnote{We use the BertForSequenceClassification provided by the HuggingFace Transformers library \citep{wolf2020huggingfaces}}. We fine-tune this model on emotion identification with the DailyDialog corpus using our canonical emotion classes. The model is trained using cross entropy loss for 4 epochs. We use the Adam optimizer \citep{kingma2014adam} with a learning rate of $2e^{-5}$ and epsilon $1e^{-8}$. We employ gradient clipping to bound the norm of gradients to 1.0. Once the BERT model is fine-tuned, we use the embedding of the [CLS]\footnote{[CLS] is a special token introduced by \citet{bert} that is used to embed an entire text sequence into a single vector.} token from the final layer as our text embedding.

\subsubsection{Training LDA-pLDA}

To train our LDA/pLDA classifier, we extract speech embeddings for all utterances in Crema-D and IEMOCAP from layers six and above of our fine-tuned speech models.  We concatenate each utterance's corresponding text embedding -- for IEMOCAP, this is straightforward.  We use the embedding for its accompanying transcript.  For Crema-D, as the spoken text is repeated across all emotions, we instead randomly sample text from DailyDialog with the same emotion label and use its embedding instead.

We train an LDA classifier to reduce the dimensionality \citep{r:beigi-sr-book-2011} of this concatenated representation to $200$ and then train a pLDA classifier \citep{ioffe2006probabilistic} on the resulting dense representations as described in 4.2.

\section{Results}

To understand the strengths of each component in our pipeline, we present the results from an exhaustive evaluation of their individual and joint predictive powers.  

\subsection{Emotion Identification}

\subsubsection{Speech TDNN}

We begin with our fine-tuned speech TDNNs whose results can be found in Table \ref{table:tdnn_results}.  We produce six different variants that vary how much learning we allow for the weights in the first six layers from our pre-trained speaker identification model, the number of layers in the fine-tuned model and the inclusion of noise augmented emotion data.  We evaluate against all the utterances in IEMOCAP with good inter-annotator agreement using the labeling presented in Table \ref{table:emotion_classification}.  

A few things are immediately clear from these results.  While noise augmentation has been shown to improve generalizability in tasks like speaker recognition by encouraging models to learn representations that are more robust, augmenting our emotion corpus with music, noise and babble degraded the performance of our TDNN.  We hypothesize that this kind of noise augmentation masks signals relevant to emotion identification.  We also find that allowing some fine-tuning of the pre-trained weights in the first six layers generally improved performance but adding an extra randomly initialized layer at the top of the network did not. It's likely that this additional layer requires more data to avoid overfitting than is typically available in an emotion corpus.

A quick comparison of our $4/5$ IEMOCAP results to the unimodal speech baselines presented in Table \ref{table:related_work} makes clear that the performance of this fine-tuned TDNN by itself is not competitive with the current state-of-the-art.  There are two likely reasons for this: 1) this is partly due to our decision to evaluate against all the utterances in IEMOCAP with the label grouping presented in Table \ref{table:emotion_classification} -- many of the baselines drop all the rare emotions that are difficult to classify; 2) the baselines transfer from pre-trained models that are much deeper and trained on much more data.  Moreover, our TDNN is just the first component in our emotion detection pipeline, meant to generate dense representations useful for subsequent in-class and out-of-class classification by pLDA.  As such, we evaluate the predictive power of pLDA classifiers trained on dense representations extracted from all six of our of TDNN variants.

\subsubsection{Speech LDA/pLDA}

In Table \ref{table:joint_plda_results}, we present the emotion identification results of several LDA/pLDA classifiers trained on speech embeddings extracted from every layer of our six TDNN variants.  We evaluate against all the utterances in IEMOCAP with the label grouping presented in \ref{table:emotion_classification}.  

Interestingly, the strongest speech-only pLDA results come from embeddings extracted from our second TDNN variant (no learning on the first size layers but with noise augmentation).  We hypothesize that, though the noise augmentation degrades the actual predictive performance of our TDNN, it produces robust embeddings in its layers that are more easily separable when classified by our pLDA model.  

Including a subset of IEMOCAP in training improves our speech-only pLDA accuracy from $0.36$ to $0.46$.  We note that while this topline performance is slightly degraded from the TDNN, the pLDA classifier is more easily adaptable to emotion classes not present in the training set.

\subsubsection{Text LDA/pLDA}

In Table \ref{table:text_plda_results}, we present the emotion identification results of an LDA/pLDA classifier trained on text embeddings extracted from our fine-tuned BERT model.  Both tasks are evaluated in the same manner as the speech pLDA classifiers.

When fine-tuned on no IEMOCAP, the performance of our text-only pLDA classifier is fairly similar to our speech-only classifier.  But, the inclusion of IEMOCAP during training dramatically boosts performance well past our speech only results (an accuracy of $0.64$).  As IEMOCAP is a scripted / acted corpus, the faithfulness of its transcripts to the emotions expressed by its utterances is high -- as a result, the text of the utterances contains a lot of signal.  Our text-based pLDA classifier is able to use this signal to make good predictions while still being easily adaptable to unseen emotion classes.

\subsubsection{Multi-modal LDA/pLDA}

In Table \ref{table:joint_plda_results}, we present the emotion identification results of LDA/pLDA classifiers trained on multi-modal concatenations of speech embeddings extracted from some of our TDNN variants and text embeddings extracted from our fine-tuned BERT model. Both tasks are evaluated in the same manner as the speech pLDA classifiers.

We first note than in both cases, without IEMOCAP and with IEMOCAP, the performance of our multi-modal pLDA classifier is improved.  This suggests that the joint representation contains some signal that wasn't present in the unimodal representations (i.e., the unimodal representations are not redundant).  While it is the case that the improvement we get with the multi-modal representation when trained on IEMOCAP is not large, the improvement we get without training on IEMOCAP (that is, evaluated against a corpus unseen during training) is meaningful.   

\begin{table}
\centering
\caption{Emotion identification results from the different variants of our fine-tuned speech TDNN.  Reported numbers are 1) "No IEMOCAP", the accuracy/F1 on IEMOCAP without training on any IEMOCAP and 2) "$4/5$ IEMOCAP", the average accuracy/F1 across the five subsets of IEMOCAP while training on the remaining four.}
\begin{tabular}[t]{lccccc}
\toprule
\multicolumn{6}{c}{TDNN: \textit{Speech Only}} \\
\midrule
\textbf{Variant} & \textbf{First Six LR} & \textbf{\# Layers} & \textbf{Noise Aug} & \textbf{No IEMOCAP} & \textbf{$4/5$ IEMOCAP} \\
\midrule
$1$ & $0$ & $7$ & $no$ & $0.408/0.269$ & $\mathbf{0.500}/\mathbf{0.418}$ \\
\vspace{0.1cm}
$2$ & $0$ & $7$ & $yes$ & $0.370/0.215$ & $0.488/0.398$ \\
\vspace{0.1cm}
$3$ & $0.0001$ & $7$ & $no$ & $\mathbf{0.426}/\mathbf{0.278}$ & $0.485/0.369$ \\
\vspace{0.1cm}
$4$ & $0.0001$ & $7$ & $yes$ & $0.322/0.186$ & $0.496/0.391$ \\
\vspace{0.1cm}
$5$ & $0.0001$ & $8$ & $no$ & $0.303/0.208$ & $0.484/0.364$ \\
\vspace{0.1cm}
$6$ & $0.0001$ & $8$ & $yes$ & $0.323/0.215$ & $0.482/0.337$ \\
\bottomrule
\end{tabular}
\label{table:tdnn_results}
\end{table}

\begin{table}
\centering
\caption{Emotion identification results from  LDA/pLDA classifiers trained on only text embeddings extracted from our fine-tuned BERT model.  Reported numbers are 1) "No IEMOCAP", the accuracy/F1/EER on IEMOCAP without training on any IEMOCAP and 2) "$4/5$ IEMOCAP", the average accuracy/F1/EER across the five subsets of IEMOCAP while training on the remaining four.}
\begin{tabular}[t]{cc}
\toprule
\multicolumn{2}{c}{pLDA: \textit{Text Only}} \\
\midrule
\multicolumn{1}{c}{\textbf{No IEMOCAP}} & \multicolumn{1}{c}{\textbf{$4/5$ IEMOCAP}} \\
\midrule
 $0.37/0.30/0.42$ & $0.64/0.59/0.29$ \\
\bottomrule
\end{tabular}
\label{table:text_plda_results}
\end{table}
\begin{table}

\centering
\caption{Emotion identification results from the different variants of our LDA/pLDA model trained on only speech embeddings extracted from different layers of our fine-tuned TDNN variants or on multi-modal concatenations of those embeddings and text embeddings extracted from our fine-tuned BERT model.  Reported numbers are 1) "No IEMOCAP", the accuracy/F1/EER on IEMOCAP without training on any IEMOCAP and 2) "$4/5$ IEMOCAP", the average accuracy/F1/EER across the five subsets of IEMOCAP while training on the remaining four.}
\begin{tabular}[t]{lcccc}
\toprule
& \multicolumn{2}{c}{pLDA: \textit{Speech Only}} & \multicolumn{2}{c}{pLDA: \textit{Speech + Text}} \\
\midrule
\multicolumn{1}{c}{\textbf{Speech Embeddings}} & \multicolumn{1}{c}{\textbf{No IEMOCAP}} & \multicolumn{1}{c}{\textbf{$4/5$ IEMOCAP}} & \multicolumn{1}{c}{\textbf{No IEMOCAP}} & \multicolumn{1}{c}{\textbf{$4/5$ IEMOCAP}} \\
\midrule
variant $2$, layer $6$ & $\mathbf{0.36}/\mathbf{0.28}/\mathbf{0.41}$ & $0.45/0.36/0.39$ & $0.34/0.26/0.41$ & - \\
\vspace{0.1cm}
variant $2$, layer $7$ & $0.36/0.23/0.42$ & $\mathbf{0.46}/\mathbf{0.36}/\mathbf{0.39}$ & $\mathbf{0.48}/\mathbf{0.37}/\mathbf{0.38}$ & $\mathbf{0.65}/\mathbf{0.59}/\mathbf{0.25}$ \\
\vspace{0.1cm}
variant $3$, layer $6$ & $0.31/0.28/0.44$ & $0.45/0.35/0.41$ & - & - \\
\vspace{0.1cm}
variant $3$, layer $7$ & $0.33/0.28/0.43$ & $0.42/0.33/0.41$ & - & - \\
\vspace{0.1cm}
variant $6$, layer $6$ & $0.34/0.29/0.43$ & $0.44/0.35/0.40$ & $0.34/0.25/0.41$ & - \\
\vspace{0.1cm}
variant $6$, layer $7$ & $0.35/0.30/0.43$ & $0.46/0.36/0.39$ & $0.41/0.34/0.38$ & $0.65/0.58/0.28$ \\
\vspace{0.1cm}
variant $6$, layer $8$ & $0.32/0.28/0.45$ & $0.44/0.35/0.40$ & $0.32/0.21/0.42$ & - \\
\bottomrule
\end{tabular}
\label{table:joint_plda_results}
\end{table}

\subsection{Emotion Confirmation}

Finally, we present \textit{emotion confirmation} results on the full IEMOCAP corpus where each pairwise test case uses the original labels in IEMOCAP (i.e., without any re-grouping) to determine whether or not two utterances are in the same class.  This emotion hypothesis testing provides a way to adapt our model to emotion classes unseen during testing, provided we have a single domain-specific example of the emotion of interest.  

As is clear from the results in Table \ref{table:emotion_confirmation}, without fine-tuning on any IEMOCAP, the best performing variant of our pLDA classifier is one that is trained on the concatenatation of speech and text embeddings.  Including $4/5$ IEMOCAP reduces this error rate from $0.457$ to $0.342$.  While these rates are admittedly high, the test scenario is a difficult one -- we include utterances whose labels were entirely unseen during training.  

\begin{table}[h]
\centering
\caption{Emotion confirmation results from  LDA/pLDA classifiers trained on on only speech embeddings extracted from different layers of our fine-tuned TDNN variants, only text embeddings extracted from our fine-tuned BERT model or on multi-modal concatenations of both.  Reported numbers are 1) "No IEMOCAP", the EER on IEMOCAP without training on any IEMOCAP and 2) "$4/5$ IEMOCAP", the average EER across the five subsets of IEMOCAP while training on the remaining four.}
\begin{tabular}[t]{lcc}
\toprule
 \multicolumn{1}{l}{\textbf{Modality}} & \multicolumn{1}{c}{\textbf{No IEMOCAP}} & \multicolumn{1}{c}{\textbf{$4/5$ IEMOCAP}} \\
\midrule
 \textit{Speech Only} & $0.457$ & $0.433$ \\
 \textit{Text Only} & $0.466$ & $0.360$ \\
 \textit{Speech + Text} & $0.457$ & $0.342$ \\
\bottomrule
\end{tabular}
\label{table:emotion_confirmation}
\end{table}

\section{Conclusion}

In this work, we present a multi-modal approach to emotion detection that first transfers learning from related tasks in speech and text to produce robust neural embeddings and then uses these embeddings to train a pLDA classifier that is able to adapt to previously unseen emotions and domains.  We show that:

\begin{enumerate}
    \item when fine-tuning on no IEMOCAP, our multi-modal pLDA classifier performs reasonably well when evaluated on the entirety of IEMOCAP (without dropping any utterances)
    \item this pLDA classifier is also able to adapt to emotions unseen during training in a one-shot classification context
    \item while our results when fine-tuning on IEMOCAP lag behind some of the larger, more data intensive networks that are the current state of the art, even pre-training a shallow model on a relatively small speaker identification corpus and then fine-tuning its weights on emotion detection provides modest gains on this task
\end{enumerate}

In the future, we think there is promise in adapting learning from such fine-tuned emotion detection models to other emotions, domains and languages via one-shot classification with pLDA.  We are also interested in exploring the effectiveness of transferring from other auxiliary tasks like automated speech recognition.

\bibliographystyle{plainnat}
\bibliography{ms}  

\begin{thebibliography}{39}
\providecommand{\natexlab}[1]{#1}
\providecommand{\url}[1]{\texttt{#1}}
\expandafter\ifx\csname urlstyle\endcsname\relax
  \providecommand{\doi}[1]{doi: #1}\else
  \providecommand{\doi}{doi: \begingroup \urlstyle{rm}\Url}\fi

\bibitem[Beigi(2011)]{r:beigi-sr-book-2011}
Homayoon Beigi.
\newblock \emph{{F}undamentals of {S}peaker {R}ecognition}.
\newblock Springer, New York, 2011.
\newblock \ ISBN: 978-0-387-77591-3,
  \url{http://www.fundamentalsofspeakerrecognition.org}.

\bibitem[Busso et~al.(2008)Busso, Bulut, Lee, Kazemzadeh, Mower, Kim, Chang,
  Lee, and Narayanan]{iemocap}
Carlos Busso, Murtaza Bulut, Chi-Chun Lee, Abe Kazemzadeh, Emily Mower, Samuel
  Kim, Jeannette~N Chang, Sungbok Lee, and Shrikanth~S Narayanan.
\newblock Iemocap: Interactive emotional dyadic motion capture database.
\newblock \emph{Language resources and evaluation}, 42\penalty0 (4):\penalty0
  335, 2008.

\bibitem[Cao et~al.(2014)Cao, Cooper, Keutmann, Gur, Nenkova, and
  Verma]{cremad}
Houwei Cao, David~G Cooper, Michael~K Keutmann, Ruben~C Gur, Ani Nenkova, and
  Ragini Verma.
\newblock Crema-d: Crowd-sourced emotional multimodal actors dataset.
\newblock \emph{IEEE transactions on affective computing}, 5\penalty0
  (4):\penalty0 377--390, 2014.

\bibitem[Chen et~al.(2019)Chen, Luo, Xu, and Ke]{compl_fusion}
Feiyang Chen, Ziqian Luo, Yanyan Xu, and Dengfeng Ke.
\newblock Complementary fusion of multi-features and multi-modalities in
  sentiment analysis.
\newblock Technical report, EasyChair, 2019.

\bibitem[Chen and Zhao(2020)]{chen2020multi}
Ming Chen and Xudong Zhao.
\newblock A multi-scale fusion framework for bimodal speech emotion
  recognition.
\newblock \emph{Proc. Interspeech 2020}, pages 374--378, 2020.

\bibitem[Chernykh and Prikhodko(2017)]{speech_1}
Vladimir Chernykh and Pavel Prikhodko.
\newblock Emotion recognition from speech with recurrent neural networks.
\newblock \emph{arXiv preprint arXiv:1701.08071}, 2017.

\bibitem[Choi et~al.(2018)Choi, Song, and Lee]{conv_atten_emo}
Woo~Yong Choi, Kyu~Ye Song, and Chan~Woo Lee.
\newblock Convolutional attention networks for multimodal emotion recognition
  from speech and text data.
\newblock In \emph{Proceedings of Grand Challenge and Workshop on Human
  Multimodal Language (Challenge-{HML})}, pages 28--34, Melbourne, Australia,
  July 2018. Association for Computational Linguistics.
\newblock \doi{10.18653/v1/W18-3304}.
\newblock URL \url{https://www.aclweb.org/anthology/W18-3304}.

\bibitem[Chung et~al.(2018)Chung, Nagrani, and Zisserman]{voxceleb}
Joon~Son Chung, Arsha Nagrani, and Andrew Zisserman.
\newblock Voxceleb2: Deep speaker recognition.
\newblock \emph{arXiv preprint arXiv:1806.05622}, 2018.

\bibitem[Devlin et~al.(2018)Devlin, Chang, Lee, and Toutanova]{bert}
Jacob Devlin, Ming-Wei Chang, Kenton Lee, and Kristina Toutanova.
\newblock Bert: Pre-training of deep bidirectional transformers for language
  understanding.
\newblock \emph{arXiv preprint arXiv:1810.04805}, 2018.

\bibitem[Dhall et~al.(2015)Dhall, Ramana~Murthy, Goecke, Joshi, and
  Gedeon]{image_1}
Abhinav Dhall, OV~Ramana~Murthy, Roland Goecke, Jyoti Joshi, and Tom Gedeon.
\newblock Video and image based emotion recognition challenges in the wild:
  Emotiw 2015.
\newblock In \emph{Proceedings of the 2015 ACM on international conference on
  multimodal interaction}, pages 423--426, 2015.

\bibitem[Fisher(1936)]{fisher1936use}
Ronald~A Fisher.
\newblock The use of multiple measurements in taxonomic problems.
\newblock \emph{Annals of eugenics}, 7\penalty0 (2):\penalty0 179--188, 1936.

\bibitem[Goel and Beigi(2020)]{goel2020cross}
Shivali Goel and Homayoon Beigi.
\newblock Cross lingual cross corpus speech emotion recognition.
\newblock \emph{arXiv preprint arXiv:2003.07996}, 2020.

\bibitem[Heusser et~al.(2019)Heusser, Freymuth, Constantin, and
  Waibel]{heusser2019bimodal}
Verena Heusser, Niklas Freymuth, Stefan Constantin, and Alex Waibel.
\newblock Bimodal speech emotion recognition using pre-trained language models.
\newblock \emph{arXiv preprint arXiv:1912.02610}, 2019.

\bibitem[Ioffe(2006)]{ioffe2006probabilistic}
Sergey Ioffe.
\newblock Probabilistic linear discriminant analysis.
\newblock In \emph{European Conference on Computer Vision}, pages 531--542.
  Springer, 2006.

\bibitem[Jack et~al.(2014)Jack, Garrod, and Schyns]{four_emotions}
Rachael~E Jack, Oliver~GB Garrod, and Philippe~G Schyns.
\newblock Dynamic facial expressions of emotion transmit an evolving hierarchy
  of signals over time.
\newblock \emph{Current biology}, 24\penalty0 (2):\penalty0 187--192, 2014.

\bibitem[Kingma and Ba(2014)]{kingma2014adam}
Diederik~P Kingma and Jimmy Ba.
\newblock Adam: A method for stochastic optimization.
\newblock \emph{arXiv preprint arXiv:1412.6980}, 2014.

\bibitem[Latif et~al.(2019)Latif, Rana, Khalifa, Jurdak, and Epps]{1904.03833}
Siddique Latif, Rajib Rana, Sara Khalifa, Raja Jurdak, and Julien Epps.
\newblock Direct modelling of speech emotion from raw speech.
\newblock \emph{arXiv preprint arXiv:1904.03833}, 2019.

\bibitem[Li et~al.(2017)Li, Su, Shen, Li, Cao, and
  Niu]{li-etal-2017-dailydialog}
Yanran Li, Hui Su, Xiaoyu Shen, Wenjie Li, Ziqiang Cao, and Shuzi Niu.
\newblock {D}aily{D}ialog: A manually labelled multi-turn dialogue dataset.
\newblock In \emph{Proceedings of the Eighth International Joint Conference on
  Natural Language Processing (Volume 1: Long Papers)}, pages 986--995, Taipei,
  Taiwan, November 2017. Asian Federation of Natural Language Processing.
\newblock URL \url{https://www.aclweb.org/anthology/I17-1099}.

\bibitem[Lian et~al.(2019)Lian, Tao, Liu, and Huang]{lian2019domain}
Zheng Lian, Jianhua Tao, Bin Liu, and Jian Huang.
\newblock Domain adversarial learning for emotion recognition.
\newblock \emph{arXiv preprint arXiv:1910.13807}, 2019.

\bibitem[Liscombe et~al.(2003)Liscombe, Venditti, and
  Hirschberg]{liscombe_features}
Jackson Liscombe, Jennifer Venditti, and Julia Hirschberg.
\newblock Classifying subject ratings of emotional speech using acoustic
  features.
\newblock In \emph{Eighth European Conference on Speech Communication and
  Technology}, 2003.

\bibitem[Mittal et~al.(2019)Mittal, Bhattacharya, Chandra, Bera, and
  Manocha]{mittal2019m3er}
Trisha Mittal, Uttaran Bhattacharya, Rohan Chandra, Aniket Bera, and Dinesh
  Manocha.
\newblock M3er: Multiplicative multimodal emotion recognition using facial,
  textual, and speech cues.
\newblock \emph{arXiv}, pages arXiv--1911, 2019.

\bibitem[Neumann and Vu(2017)]{speech_2}
Michael Neumann and Ngoc~Thang Vu.
\newblock Attentive convolutional neural network based speech emotion
  recognition: A study on the impact of input features, signal length, and
  acted speech.
\newblock \emph{arXiv preprint arXiv:1706.00612}, 2017.

\bibitem[Pappagari et~al.(2020)Pappagari, Wang, Villalba, Chen, and
  Dehak]{pappagari2020x}
Raghavendra Pappagari, Tianzi Wang, Jesus Villalba, Nanxin Chen, and Najim
  Dehak.
\newblock x-vectors meet emotions: A study on dependencies between emotion and
  speaker recognition.
\newblock In \emph{ICASSP 2020-2020 IEEE International Conference on Acoustics,
  Speech and Signal Processing (ICASSP)}, pages 7169--7173. IEEE, 2020.

\bibitem[Poria et~al.(2019)Poria, Hazarika, Majumder, Naik, Cambria, and
  Mihalcea]{meld}
Soujanya Poria, Devamanyu Hazarika, Navonil Majumder, Gautam Naik, Erik
  Cambria, and Rada Mihalcea.
\newblock Meld: A multimodal multi-party dataset for emotion recognition in
  conversations.
\newblock In \emph{Proceedings of the 57th Annual Meeting of the Association
  for Computational Linguistics}, pages 527--536, 2019.

\bibitem[Povey et~al.(2011)Povey, Ghoshal, Boulianne, Burget, Glembek, Goel,
  Hannemann, Motlicek, Qian, Schwarz, et~al.]{povey2011kaldi}
Daniel Povey, Arnab Ghoshal, Gilles Boulianne, Lukas Burget, Ondrej Glembek,
  Nagendra Goel, Mirko Hannemann, Petr Motlicek, Yanmin Qian, Petr Schwarz,
  et~al.
\newblock The kaldi speech recognition toolkit.
\newblock In \emph{IEEE 2011 workshop on automatic speech recognition and
  understanding}, number CONF. IEEE Signal Processing Society, 2011.

\bibitem[Rana(2016)]{bipolar}
Rajib Rana.
\newblock Poster: Context-driven mood mining.
\newblock In \emph{Proceedings of the 14th Annual International Conference on
  Mobile Systems, Applications, and Services Companion}, pages 143--143, 2016.

\bibitem[Rana et~al.(2019)Rana, Latif, Gururajan, Gray, Mackenzie, Humphris,
  and Dunn]{distress}
Rajib Rana, Siddique Latif, Raj Gururajan, Anthony Gray, Geraldine Mackenzie,
  Gerald Humphris, and Jeff Dunn.
\newblock Automated screening for distress: A perspective for the future.
\newblock \emph{European journal of cancer care}, 28\penalty0 (4):\penalty0
  e13033, 2019.

\bibitem[Scherer and Ekman(2014)]{ekman}
Klaus~R Scherer and Paul Ekman.
\newblock \emph{Approaches to emotion}.
\newblock Psychology Press, 2014.

\bibitem[Snyder et~al.(2018)Snyder, Garcia-Romero, Sell, Povey, and
  Khudanpur]{xvectors}
David Snyder, Daniel Garcia-Romero, Gregory Sell, Daniel Povey, and Sanjeev
  Khudanpur.
\newblock X-vectors: Robust dnn embeddings for speaker recognition.
\newblock In \emph{2018 IEEE International Conference on Acoustics, Speech and
  Signal Processing (ICASSP)}, pages 5329--5333. IEEE, 2018.

\bibitem[Tripathi et~al.(2018{\natexlab{a}})Tripathi, Tripathi, and
  Beigi]{beigi_iemocap}
Samarth Tripathi, Sarthak Tripathi, and Homayoon Beigi.
\newblock Multi-modal emotion recognition on iemocap dataset using deep
  learning.
\newblock \emph{arXiv preprint arXiv:1804.05788}, 2018{\natexlab{a}}.

\bibitem[Tripathi et~al.(2018{\natexlab{b}})Tripathi, Tripathi, and
  Beigi]{tripathi2018multi}
Samarth Tripathi, Sarthak Tripathi, and Homayoon Beigi.
\newblock Multi-modal emotion recognition on iemocap dataset using deep
  learning.
\newblock \emph{arXiv preprint arXiv:1804.05788}, 2018{\natexlab{b}}.

\bibitem[Tripathi et~al.(2019)Tripathi, Kumar, Ramesh, Singh, and
  Yenigalla]{tripathi2019deep}
Suraj Tripathi, Abhay Kumar, Abhiram Ramesh, Chirag Singh, and Promod
  Yenigalla.
\newblock Deep learning based emotion recognition system using speech features
  and transcriptions.
\newblock \emph{arXiv preprint arXiv:1906.05681}, 2019.

\bibitem[Tzirakis et~al.(2017)Tzirakis, Trigeorgis, Nicolaou, Schuller, and
  Zafeiriou]{1704.08619}
Panagiotis Tzirakis, George Trigeorgis, Mihalis~A Nicolaou, Bj{\"o}rn~W
  Schuller, and Stefanos Zafeiriou.
\newblock End-to-end multimodal emotion recognition using deep neural networks.
\newblock \emph{IEEE Journal of Selected Topics in Signal Processing},
  11\penalty0 (8):\penalty0 1301--1309, 2017.

\bibitem[Waibel et~al.(1989)Waibel, Hanazawa, Hinton, Shikano, and
  Lang]{waibel1989phoneme}
Alex Waibel, Toshiyuki Hanazawa, Geoffrey Hinton, Kiyohiro Shikano, and Kevin~J
  Lang.
\newblock Phoneme recognition using time-delay neural networks.
\newblock \emph{IEEE transactions on acoustics, speech, and signal processing},
  37\penalty0 (3):\penalty0 328--339, 1989.

\bibitem[Wolf et~al.(2020)Wolf, Debut, Sanh, Chaumond, Delangue, Moi, Cistac,
  Rault, Louf, Funtowicz, Davison, Shleifer, von Platen, Ma, Jernite, Plu, Xu,
  Scao, Gugger, Drame, Lhoest, and Rush]{wolf2020huggingfaces}
Thomas Wolf, Lysandre Debut, Victor Sanh, Julien Chaumond, Clement Delangue,
  Anthony Moi, Pierric Cistac, Tim Rault, Rémi Louf, Morgan Funtowicz, Joe
  Davison, Sam Shleifer, Patrick von Platen, Clara Ma, Yacine Jernite, Julien
  Plu, Canwen Xu, Teven~Le Scao, Sylvain Gugger, Mariama Drame, Quentin Lhoest,
  and Alexander~M. Rush.
\newblock Huggingface's transformers: State-of-the-art natural language
  processing, 2020.

\bibitem[Yoon et~al.(2018)Yoon, Byun, and Jung]{1810.04635}
Seunghyun Yoon, Seokhyun Byun, and Kyomin Jung.
\newblock Multimodal speech emotion recognition using audio and text.
\newblock In \emph{2018 IEEE Spoken Language Technology Workshop (SLT)}, pages
  112--118. IEEE, 2018.

\bibitem[Yoon et~al.(2020)Yoon, Dey, Lee, and Jung]{yoon2020attentive}
Seunghyun Yoon, Subhadeep Dey, Hwanhee Lee, and Kyomin Jung.
\newblock Attentive modality hopping mechanism for speech emotion recognition.
\newblock In \emph{ICASSP 2020-2020 IEEE International Conference on Acoustics,
  Speech and Signal Processing (ICASSP)}, pages 3362--3366. IEEE, 2020.

\bibitem[Zhou and Beigi(2020)]{zhou2020transfer}
Sitong Zhou and Homayoon Beigi.
\newblock A transfer learning method for speech emotion recognition from
  automatic speech recognition.
\newblock \emph{arXiv preprint arXiv:2008.02863}, 2020.

\bibitem[Zhu et~al.(2017)Zhu, Shang, Shao, and Guo]{depression}
Yu~Zhu, Yuanyuan Shang, Zhuhong Shao, and Guodong Guo.
\newblock Automated depression diagnosis based on deep networks to encode
  facial appearance and dynamics.
\newblock \emph{IEEE Transactions on Affective Computing}, 9\penalty0
  (4):\penalty0 578--584, 2017.

\end{thebibliography}

\end{document}